\documentclass[11pt,a4paper]{article}

\usepackage{ellipsis}
\usepackage[final,babel]{microtype}

\usepackage{jcappub}

\usepackage{lmodern}
\usepackage[T1]{fontenc}
\usepackage[utf8]{inputenc}
\usepackage[scaled=0.95]{inconsolata}

\usepackage{enumitem}
\usepackage[english]{babel}

\usepackage{booktabs}
\usepackage{hyperref}
\usepackage{amsmath}
\usepackage{subcaption}
\usepackage{listings}
\usepackage{pythonhighlight}

\newcommand{\mean}[1]{\left\langle #1 \right\rangle}
\newcommand{\largemean}[1]{\big\langle #1 \big\rangle}
\newcommand{\like}{\mathcal{L}}
\newcommand{\gev}{\,\text{GeV}}
\newcommand{\cmsq}{\,\text{cm}^2}
\newcommand{\kms}{\,\text{km/s}}
\newcommand{\kg}{\,\text{kg}}
\newcommand{\sigsip}{\sigma}
\newcommand{\refsec}{Sec.~\ref}
\newcommand{\refeq}{Eq.~\ref}
\newcommand{\refcite}{Ref.~\cite}
\newcommand{\refapp}{App.~\ref}
\newcommand{\reffig}{Fig.~\ref}
\newcommand{\refsubfig}[1]{(\subref{#1})}
\let\vec\boldsymbol
\newcommand{\dif}{\,d}

\newcommand{\pg}[2]{p\left(#1\,\boldsymbol{|}\,#2\right)}
\newcommand{\Pg}[2]{P\left(#1\,\boldsymbol{|}\,#2\right)}
\renewcommand{\P}[1]{P\left(#1\right)}
\newcommand{\stirling}{\genfrac\{\}{0pt}{}}
\renewcommand{\binom}[2]{\left(\genfrac{}{}{0pt}{}{#1}{#2}\right)}

\usepackage{xspace}
\newcommand{\xenon}{XENON1T\xspace}

\setlist[description]{font=\boldmath\bfseries\textbullet\space}

\title{Non-parametric uncertainties in the dark matter velocity distribution}
\author{Andrew Fowlie}
\affiliation{Department of Physics and Institute of Theoretical Physics, Nanjing Normal University, Nanjing, Jiangsu 210023, China}
\affiliation{School of Physics and Astronomy, Monash University, Melbourne, Australia}
\emailAdd{andrew.j.fowlie@qq.com}
\notoc
\abstract{We investigate the impact of uncertainty in the velocity distribution of dark matter on direct detection experiments. We construct an multinomial prior with a hyperparameter $\beta$ that describes the strength of our belief in an isotropic Maxwell-Boltzmann velocity distribution. By varying $\beta$, we interpolate between a halo-independent and halo-dependent analysis. We present a novel approximation for the marginalisation of this prior that is applicable to any counting experiment. With this formula, we investigate the impact of the uncertainty in limits from \xenon. For dark matter masses greater than about $60\gev$, we find extremely mild sensitivity to the distribution. Below about $60\gev$, the limit weakens by less than an order of magnitude if we assume an isotropic distribution in the galactic frame. If we permit anisotropic distributions, the limit further weakens, but at most by about two orders of magnitude. Lastly, we check the impact of parametric uncertainties and discuss the possible inclusion and impact of our technique in global fits.
}

\begin{document}

\maketitle

\section{Introduction}

There is evidence from gravitational interactions for the existence of dark matter (DM) throughout our Universe (see e.g., \refcite{Bertone:2004pz}). Weakly interacting massive particles (WIMPs) are a popular candidate for DM, since they naturally arise in many well-motivated extensions of the Standard Model (SM), e.g., supersymmetry, and correctly predict the relic abundance of DM by the so-called WIMP miracle~\cite{Jungman:1995df}. As WIMPs must annihilate to SM particles in the early Universe, by crossing symmetry, we expect that WIMPs must scatter elastically with SM particles. No evidence of such scattering was found in direct detection (DD) experiments by, inter alia, \xenon~\cite{Aprile:2018dbl}, LUX~\cite{Akerib:2016lao} or PandaX~\cite{Cui:2017nnn}, resulting in upper limits on the DM scattering cross section with nucleons, e.g., for a $35\gev$ DM particle the cross section must be less than about $10^{-46}\cmsq$~\cite{Aprile:2018dbl}. The limits depend upon assumptions about the velocity distribution of DM. The correct treatment and impact of uncertainties in the velocity distribution are the subjects of this work.

From only theoretical considerations we anticipate that the velocity distribution could be similar to a Maxwell-Boltzmann distribution,
\begin{equation}\label{eq:mb}
m(v, \cos\theta, \phi) \propto \begin{cases}
        v^2 e^{-\left(v / v_0\right)^2} & v < v_\text{esc}\\
        0 & v \ge v_\text{esc}
        \end{cases},
\end{equation}
where  $v_\text{esc}$ and $v_0$ are the escape and modal velocities, respectively, and $\int_{\vec v} m(\vec v) d^3v = 1$. Indeed, this is the distribution that is assumed by DD experiments, including \xenon. There are parametric uncertainties in the escape and modal velocities. There are, furthermore, non-parametric uncertainties as we know that departures from Maxwell-Boltzmann are plausible (and in fact may be preferred; see e.g., \refcite{Necib:2018iwb,Mandal:2018efq}). A Maxwell-Boltzmann follows from assuming a spherically-symmetric, isothermal halo of collisionless DM particles with density $\rho(r) \propto 1/r^2$; each assumption is questionable (see e.g.,~\refcite{Ibarra:2017mzt}). We recently proposed a non-parametric treatment of this state of knowledge~\cite{Fowlie:2017ufs} using the formalism of quantified maximum entropy (QME). We did not assume any particular parametric distribution for the velocity distribution; instead, we constructed an entropic prior for the velocity distribution that peaked at a Maxwell-Boltzmann and penalised departures from Maxwell-Boltzmann according to the relative entropy,
\begin{equation}\label{eq:entropy}
S[f, m] = - \int_{\vec v} f(\vec v) \ln \left(\frac{f(\vec v)}{m(\vec v)}\right) \dif^3v.
\end{equation}
The techniques in \refcite{Fowlie:2017ufs} relied on a Laplace approximation and were difficult to apply. A further drawback of QME is that the results depend upon details of the discretization of the velocity and that in the continuum limit it suffers from the law of large numbers, such that it overwhelming favours a Maxwell-Boltzmann (see \refsec{sec:recap}).

In this work we present a similar entropic prior that overcomes this drawback. In \refsec{sec:dd}, we review our treatment of the expected number of signal events in a DD experiment. In \refsec{sec:recap}, we recapitulate the pertinent aspects of \refcite{Fowlie:2017ufs} and the merits of our use of the relative entropy, before presenting a formula for the marginalisation of an entropic prior based on a multinomial process. In \refsec{sec:numeric} we apply it to recent results from \xenon (2018). Lastly,
we conclude in
\refsec{sec:conclusions}. We furthermore motivate and discuss our new formula in \refapp{app:average} and \refapp{app:proof}, respectively, and present our code that implements it in \refapp{app:code}.

\section{Events function}\label{sec:dd}

The number of expected events, $\lambda$, in a DD experiment such as \xenon may be expressed as an expectation of the velocity distribution in the galactic frame, $f$,
\begin{equation}\label{eq:lambda}
\lambda = \mean{w}_f \equiv \int f(\vec v) \cdot w(\vec v) \dif^3 v.
\end{equation}
where $\mean{y}_f \equiv \int y(\vec v) \cdot f(\vec v) \dif^3 v$ indicates an average over the velocity distribution, $f$, and the function $w(\vec v)$ defines the number of expected events as a function of the DM velocity in the galactic frame, $\vec v$.
We define it in the laboratory frame and transform it to the galactic frame by a Galilean boost. In the laboratory frame it is isotropic and may be written as
\begin{equation}
w_\text{lab}(v) = \frac{2 MT \rho}{m_\chi} \cdot v \cdot \int \frac{\dif\sigma}{\dif q^2} \cdot \Phi(E) \dif E + b,
\end{equation}
where $MT$ is the exposure; $\Phi(E)$ is the detector efficiency at recoil energy $E$; $b$ is the expected number of background events; $m_\chi$ and $\rho$ are the mass and local density of DM, respectively; and ${\dif\sigma}/{\dif q^2}$ is the differential cross section. We assume that the interactions are velocity and momentum independent such that the differential cross section may be written as
\begin{equation}
\frac{\dif\sigma}{\dif q^2} = \frac{\sigsip}{4\mu^2 v^2} \cdot F^2(q) \cdot \theta(q_\text{max} -  q),
\end{equation}
where $\sigma$ is the scattering cross section at zero momentum; the momentum $q^2 = 2 m_n E$; by kinematics $q_\text{max} = 2 \mu v$; $\mu$ is the reduced mass of the DM and nucleon; $m_n$ is the nucleon mass; $F$ is a nuclear form-factor; and $\theta$ denotes a stepfunction.

Our treatment of the expected number of events differs from the canonical one (see e.g., \refcite{Workgroup:2017lvb}) only in our presentation; we reversed the order of the energy and velocity integrals and boosted $w(\vec v)$ to the galactic frame rather than $f(\vec v)$ to the laboratory frame. This approach was introduced in \refcite{DelNobile:2013cva,Gondolo:2017jro}.

\section{Entropic prior}\label{sec:recap}

We recently proposed treating uncertainties in the velocity distribution with quantified maximum entropy~\cite{Fowlie:2017ufs}. Rather than assuming any
particular velocity distribution, we constructed a prior upon possible velocity distributions and averaged upon it. The prior penalised departures from a default distribution by the relative entropy,
\begin{equation}\label{eq:QME}
\pg{\vec f}{\vec m} \propto \frac{e^{\beta S[f, m]}}{\prod_i \sqrt{f_i}}  \cdot \delta\left(\sum f_i - 1\right).
\end{equation}
where $S[f, m]$, defined in \refeq{eq:entropy}, is the entropy of the velocity distribution, $f$, relative to a Maxwellian, $m$, and we denote a
discrete distribution across $r$ bins by $\vec f = \{f_1, f_2, \dots, f_r\}$, and similarly for the default distribution $\vec m$. When $f = m$, the entropy vanishes, and it is otherwise negative. The hyperparameter $\beta$ represented the strength of our conviction that the velocity distribution is Maxwellian; as $\beta\to \infty$, our uncertainty vanished and the prior selected $f = m$, and as $\beta\to0$, the penalty for departures from the default model, $m$, vanished. Thus by varying $\beta$, we interpolated between a halo-independent ($\beta\to 0$) and halo-dependent ($\beta\to\infty$) approach.

Following \refcite{Fowlie:2017ufs}, a frequentist treatment was proposed~\cite{Ibarra:2018yxq} in which departures were measured by
\begin{equation}
\Delta[f, m] = \max_{\vec v} \left|\frac{f(\vec v) - m(\vec v)}{m(\vec v)}\right|.
\end{equation}
Rather than averaging upon a set of velocity distributions, a distribution was found that maximised the likelihood subject to an upper bound on the discrepancy, $\Delta[f, m]$. There were two main advantages to our approach. First, there is an information theoretic meaning to the relative entropy and the entropic prior may be derived as a unique choice subject to modest axioms (see e.g., \refcite{Skilling1988,Skilling1989}), whereas $\Delta[f, m]$ is ad hoc. Second, we coherently incorporated uncertainty by marginalising rather than profiling. Only the former respects the fact that the plausibilities of disjoint propositions should sum.

We note, however, a subtle drawback in the QME prior: the results are sensitive to the discretization of the velocity and are counter-intuitive in the continuum limit~\cite{doi:10.1111/1467-9868.00065}. In that limit there are an infinite number of contributions to the probability in a macroscopic interval, $\Delta v$, i.e., in the continuum limit, the sum,
\begin{equation}
    f_i \equiv \P{v \le v^\prime \le v + \Delta v} = \sum_{i=1}^n \P{v + \tfrac{(i - 1)\Delta v}{n}  \le v^\prime \le v + \tfrac{i \Delta v}{n}},
\end{equation}
contains an infinite number of terms as $n\to\infty$. By the law of large numbers, for the QME prior the probability in such an interval equals its expected value, $f_i \to m_i$. This means that the QME prior overwhelming favours the default distribution on macroscopic scales. There is thus a delicate interplay between $\beta$ and the discretization of the velocity; although $\beta$ penalises departures from the default distribution, in the continuum limit it operates at the microscopic scale, $dv$. On macroscopic scales, by the law of large numbers, departures average away.

We could avoid this problem by specifying a default distribution (e.g., a Maxwellian) and a finite scale $\Delta v$ below which we wish to penalise departures from it. We instead avoid it by generating velocity distributions by scattering $\beta$ quanta of probability on possible velocities. We initially discretize the velocity distribution, $f_i \approx f(\vec v_i) \Delta v^3$, by dividing the velocity into $r$ bins of volume $\Delta v^3$, but ultimately we take a continuum limit. We assume that the quanta fall into particular bins with  probabilities from the default model, $m_i$. This is a multinomial process. The law of large numbers strikes only in the limit $\beta\to\infty$, forcing the velocity distribution to the default one, as desired. This is detailed in \refapp{app:average}.
This choice is motivated by the fact that just like the QME prior, we find that our prior penalizes departures from a parametric model according to the relative entropy,
\begin{equation}
\Pg{\vec f}{\vec m} \propto e^{\beta S[\vec f, \vec m]}.
\end{equation}
In fact, our prior approximates the QME one when the number of bins in the QME prior $r \lesssim \beta$, which implies a bin width $\Delta v \gtrsim v_\text{esc} / \beta$. It differs from the QME prior in that it requires $f_i$ to be quantized in multiples of $1/\beta$.

To incorporate uncertainty in the velocity distribution, we begin from a Poisson probability for observing $q$ events given that $\lambda$ events were expected,
\begin{equation}\label{eq:poisson}
\like \equiv \Pg{q}{\lambda} = \frac{e^{-\lambda} \lambda^q}{q!}.
\end{equation}
We note that the expected number of events is a function of the DM mass, scattering cross section with nucleons and velocity distribution, i.e, $\lambda \equiv \lambda(m_\chi, \sigsip, \vec f)$.
We want to marginalise upon the velocity distribution, i.e., calculate the sum,
\begin{equation}\label{eq:integral}
\mean{\like} \equiv \Pg{q}{m_\chi, \sigsip} = \sum \Pg{q}{m_\chi, \sigsip, \vec f} \cdot \Pg{\vec f}{\vec m} = \sum \frac{e^{-\lambda} \lambda^q}{q!} \cdot \Pg{\vec f}{\vec m},
\end{equation}
where we marginalised the velocity distribution over our prior, $\Pg{\vec f}{\vec m}$.
We compute the sum exactly in \refapp{app:proof}. For the experiment that we investigate, \xenon (2018), the number of observed events  was $q=2$ such that using \refeq{eq:result} we find,
\begin{equation}\label{eq:average_like}
\mean{\like} =
\frac{1}{2}
\mean{e^{-w / \beta}}_m^\beta
\left(
\frac{\beta - 1}{\beta} \frac{\largemean{w e^{-w / \beta}}_m^2}{\largemean{e^{-w / \beta}}_m^2} + \frac{1}{\beta} \frac{\largemean{w^2 e^{-w / \beta}}_m}{\largemean{e^{-w / \beta}}_m}
\right),
\end{equation}
for integer  $\beta\ge 1$ and where $\mean{y}_m \equiv \int y(\vec v) \cdot m(\vec v) \dif^3 v$ indicates an average over the default model, i.e., a Maxwell-Boltzmann. For the common case in which no events were observed, $q=0$, we find,
\begin{equation}
\mean{\like} = \mean{e^{-w / \beta}}_m^\beta.
\end{equation}
The marginalized likelihoods resemble our original Poisson likelihood in \refeq{eq:poisson}, which in a similar notation for $q=2$ reads
\begin{equation}\label{eq:beta_infinity}
\like = \frac12 e^{-\mean{w}_m} \mean{w}_m^2.
\end{equation}
The changes result from our
incorporation of the uncertainty in the velocity distribution. In the limit in which our uncertainty vanishes, $\beta\to\infty$, we indeed recover \refeq{eq:beta_infinity},
\begin{equation}
\lim_{\beta\to\infty} \mean{\like} = \like.
\end{equation}
We thus interpret our treatment as a non-parametric relaxation of a default distribution. We cannot, however, throw away all information about the default model as our multinomial process requires $\beta \ge 1$.

\subsection{Isotropy}\label{sec:isotropy}

Our averaged likelihood in \refeq{eq:average_like} makes no assumptions about isotropy --- it averages over anisotropic and isotropic velocity distributions weighted by an entropic prior. If we wish to assume isotropy, we must pick an isotropic default model (such as the Maxwell-Boltzmann),
\begin{equation}
m(v, \cos\theta, \phi) = \frac{1}{4\pi} m(v) ,
\end{equation}
and omit an entropic prior for the angular variables, $m(\cos\theta, \phi)$, such that there are no deviations from isotropy. In our formalism, this is mathematically equivalent to replacing the event function by
\begin{equation}
w(\vec v) \to w(v) = {1}/{4\pi}\int w(\vec v) \dif\cos\theta \dif\phi
\end{equation}
throughout, i.e., using an angle-averaged event function.

\section{Impact on \xenon (2018) limits}\label{sec:numeric}

\subsection{Events function}\label{sec:events}

\begin{figure}[t]
    \centering
    \begin{subfigure}[t]{0.48\textwidth}
        \centering
        \includegraphics[height=0.26\textheight]{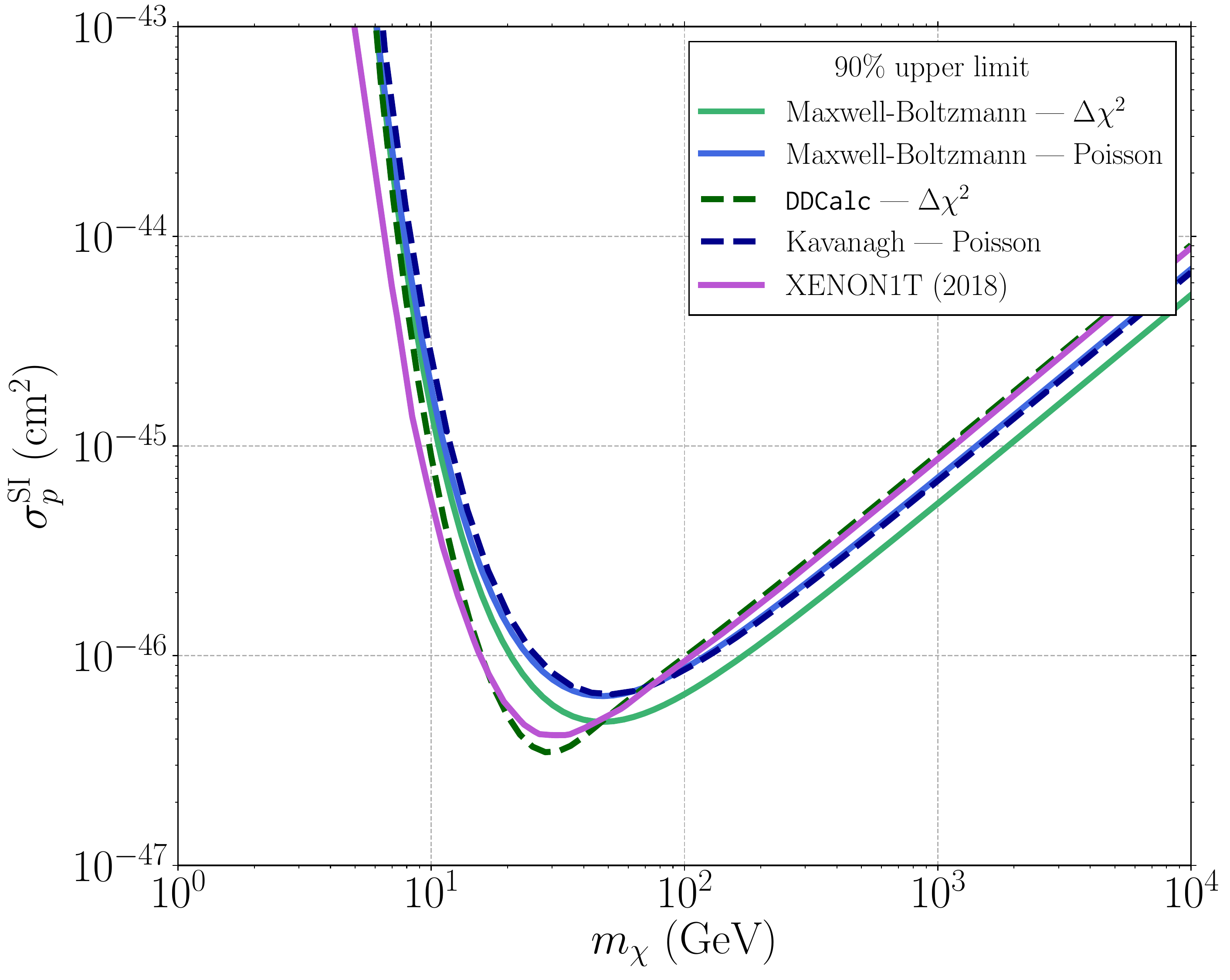}
        \caption{The \xenon $90\%$ bound}
        \label{fig:limit}
    \end{subfigure}
    \begin{subfigure}[t]{0.48\textwidth}
        \centering
        \includegraphics[height=0.26\textheight]{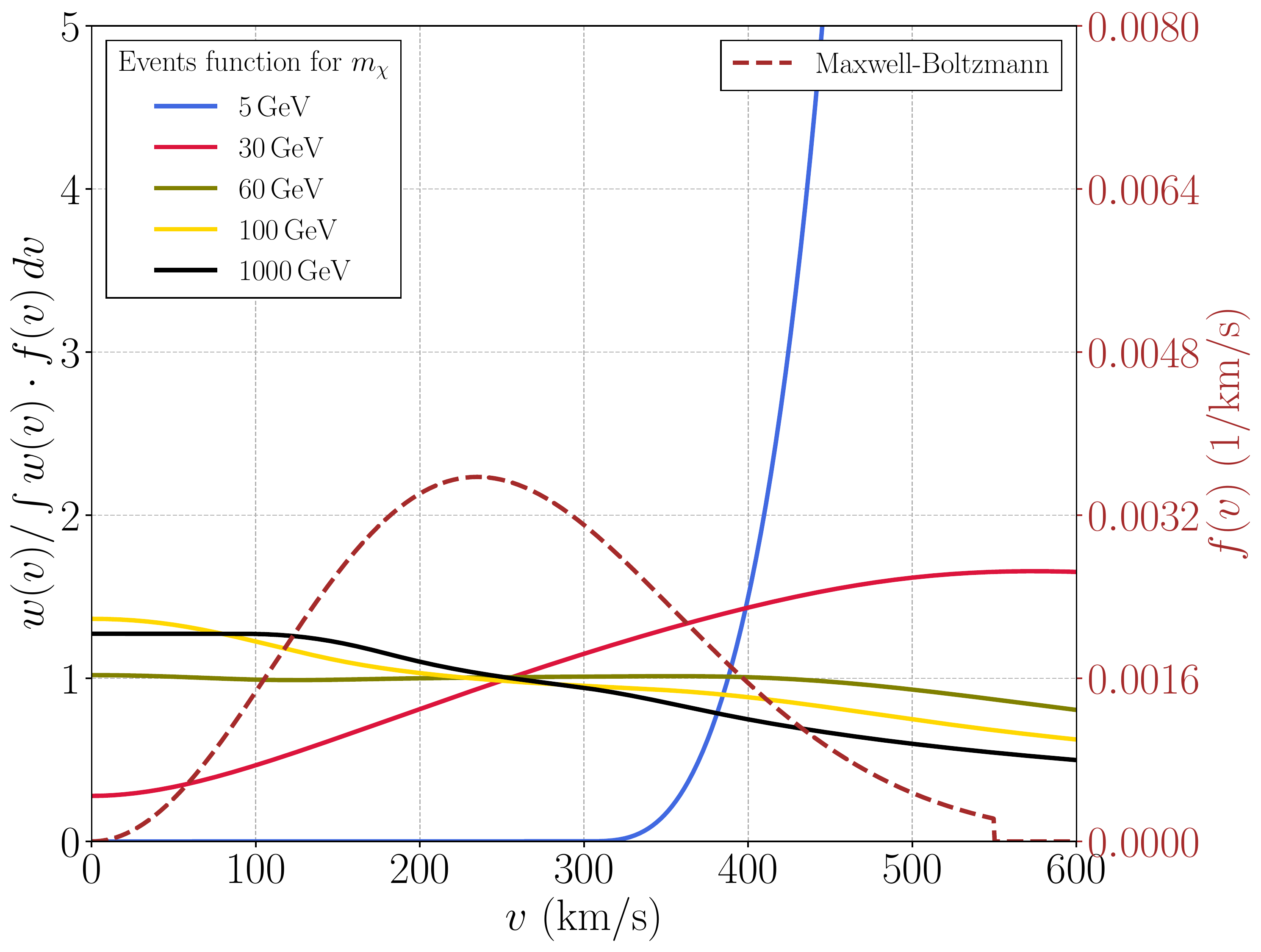}
        \caption{Angle-averaged events function}
        \label{fig:events}
    \end{subfigure}
    \caption{Validation of our events function, $w(\vec v)$. In \refsubfig{fig:limit} we compare the \xenon $90\%$ bound (solid violet); our reproductions from $\Delta\chi^2$ (solid green) and Poisson statistics (solid blue); and reproductions from \refcite{Workgroup:2017lvb,Athron:2018hpc} (dashed green) and \refcite{github} (dashed blue). In \refsubfig{fig:events} we show the angle-averaged number of events per velocity, $w(v)$, normalised to the number of expected events for five different DM masses (solid lines). For reference we show a Maxwell-Boltzmann distribution (dashed brown) with a separate vertical axis.}\label{fig:reproduction}
\end{figure}

To validate our treatment of the \xenon (2018) experiment, we first reproduced the \xenon (2018) $90\%$ upper bound on the spin-independent scattering cross section with nucleons assuming a Maxwell-Boltzmann velocity distribution. With \refeq{eq:poisson} we calculated the likelihood of $2$ events in $278.8$ days, given a DM signal with a particular mass and cross section, and $1.62$ expected background events. We considered only events in the reference region of the $900\kg$ inner detector, $M = 0.475 \times 900\kg$.\footnote{See Tab.~1 of \refcite{Aprile:2018dbl}.}

We show our results in \reffig{fig:limit}. We calculated a $90\%$ limit at $3.7$ signal events from Poisson statistics, which closely matches a similar reproduction~\cite{github}. Our $90\%$ limit from Wilks' theorem is similar to that from \texttt{DDCalc}~\cite{Workgroup:2017lvb,Athron:2018hpc}, which used additional binning information but in only the $650\kg$ inner detector. The minor differences between our limit and the \xenon one were expected since \xenon used spectral information and an unbinned analysis. Thus we are satisfied that we successfully computed the events function for the \xenon experiment, $w(\vec v)$, which appears in our treatment of the uncertainty in the velocity distribution.

We plot our angle-averaged events function for five different DM masses in \reffig{fig:events}. We find that for lower DM masses, as expected from kinematics, the events function peaks at higher velocity. This explains the loss in sensitivity for light DM masses: for light DM masses, the signal vanishes as the events functions favours high velocities but the velocity distribution is zero beyond the escape velocity. For higher DM masses, the sensitivity deteriorates as the number density, $\rho / m_\chi$, shrinks as the DM mass increases. We note that for $m_\chi \simeq 60 \gev$ the events function is particularly flat.

\subsection{Isotropic velocity distribution}

\begin{figure}[t]
    \centering
    \begin{subfigure}[t]{0.48\textwidth}
        \centering
        \includegraphics[height=0.26\textheight]{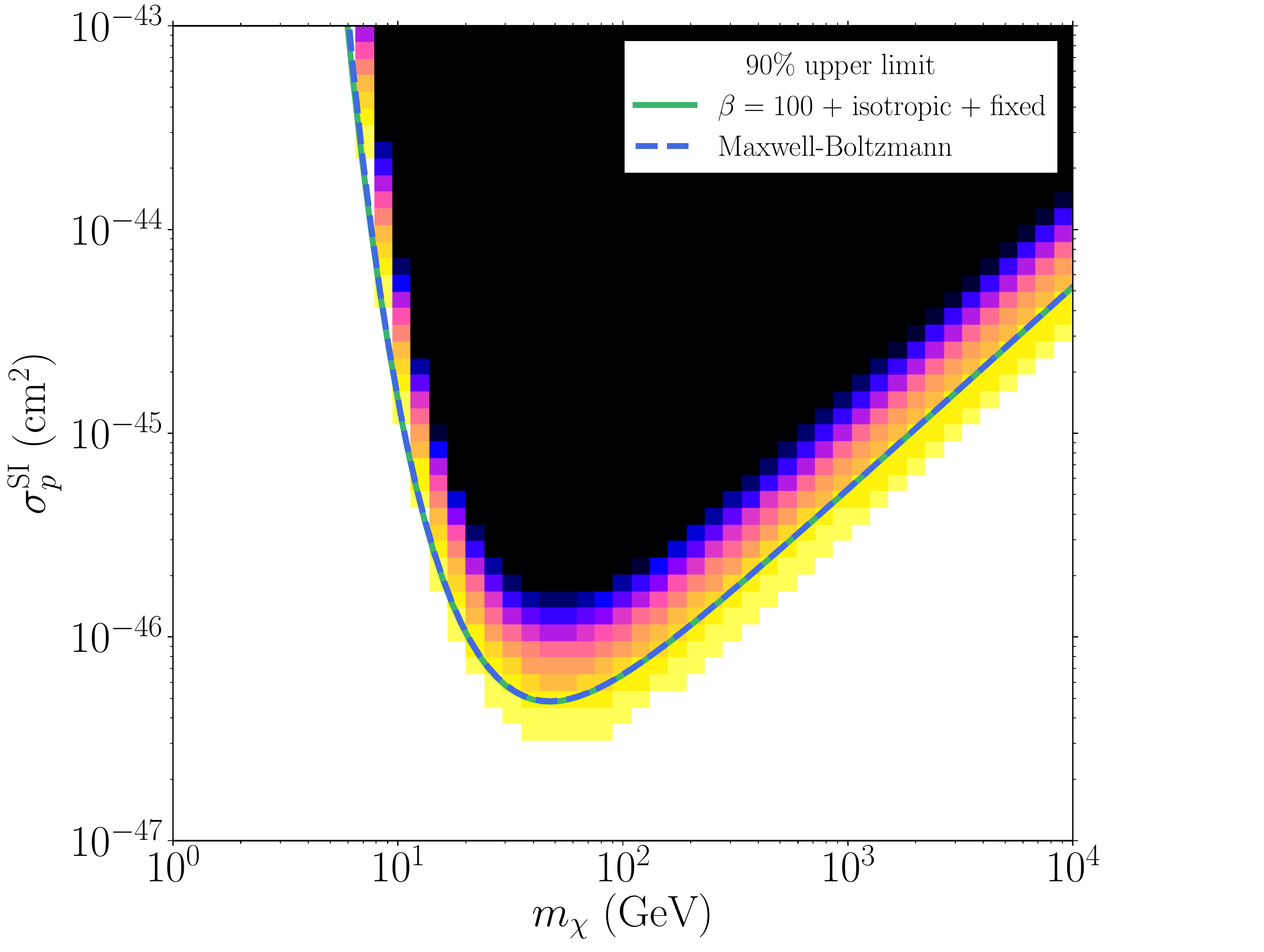}
        \caption{$\beta = 100$}
        \label{fig:isotropic_a}
    \end{subfigure}
    \hspace{-0.75cm}
    \begin{subfigure}[t]{0.48\textwidth}
        \centering
        \includegraphics[height=0.26\textheight]{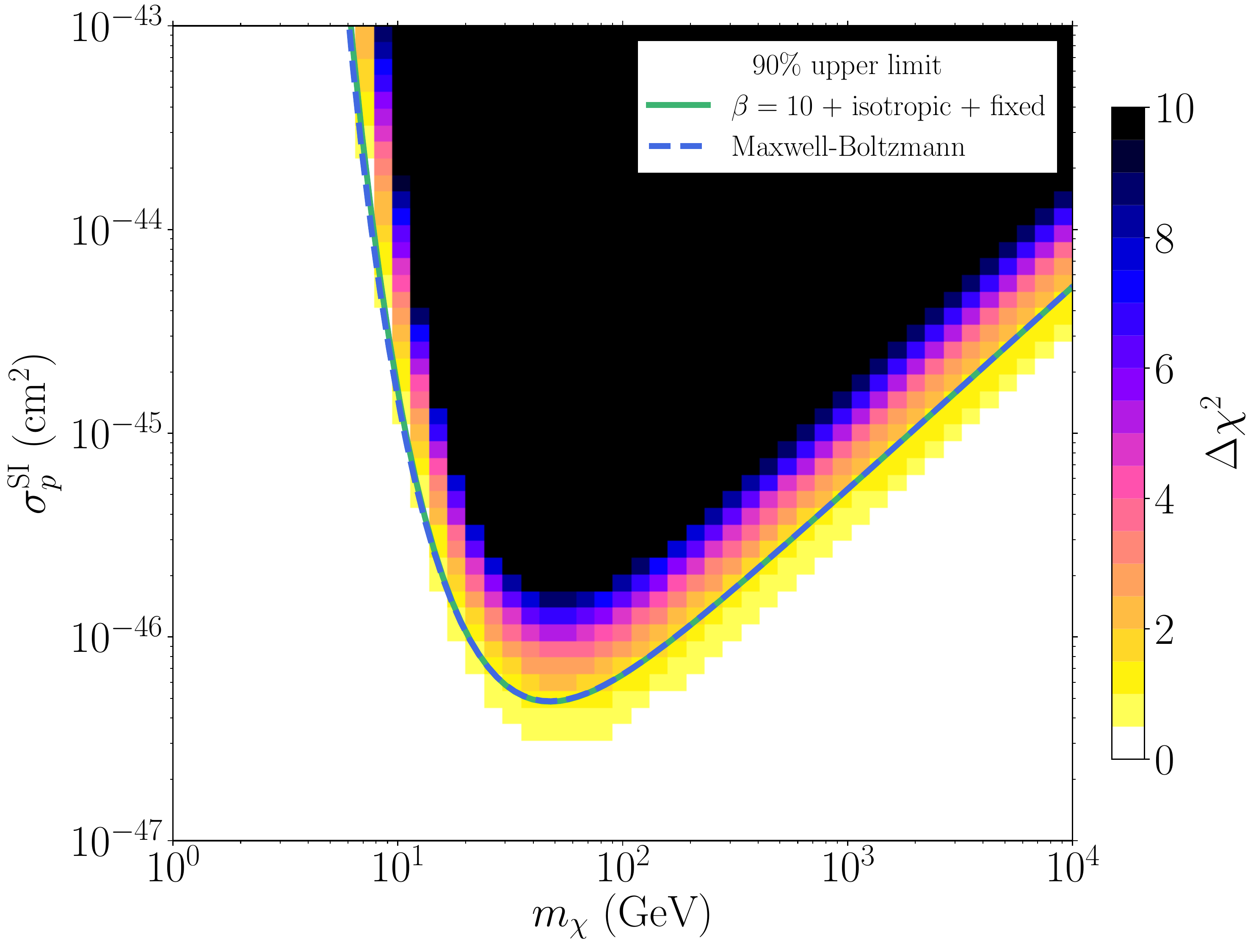}
        \caption{$\beta = 10$}
        \label{fig:isotropic_b}
    \end{subfigure}

    \begin{subfigure}[t]{0.48\textwidth}
        \centering
        \includegraphics[height=0.26\textheight]{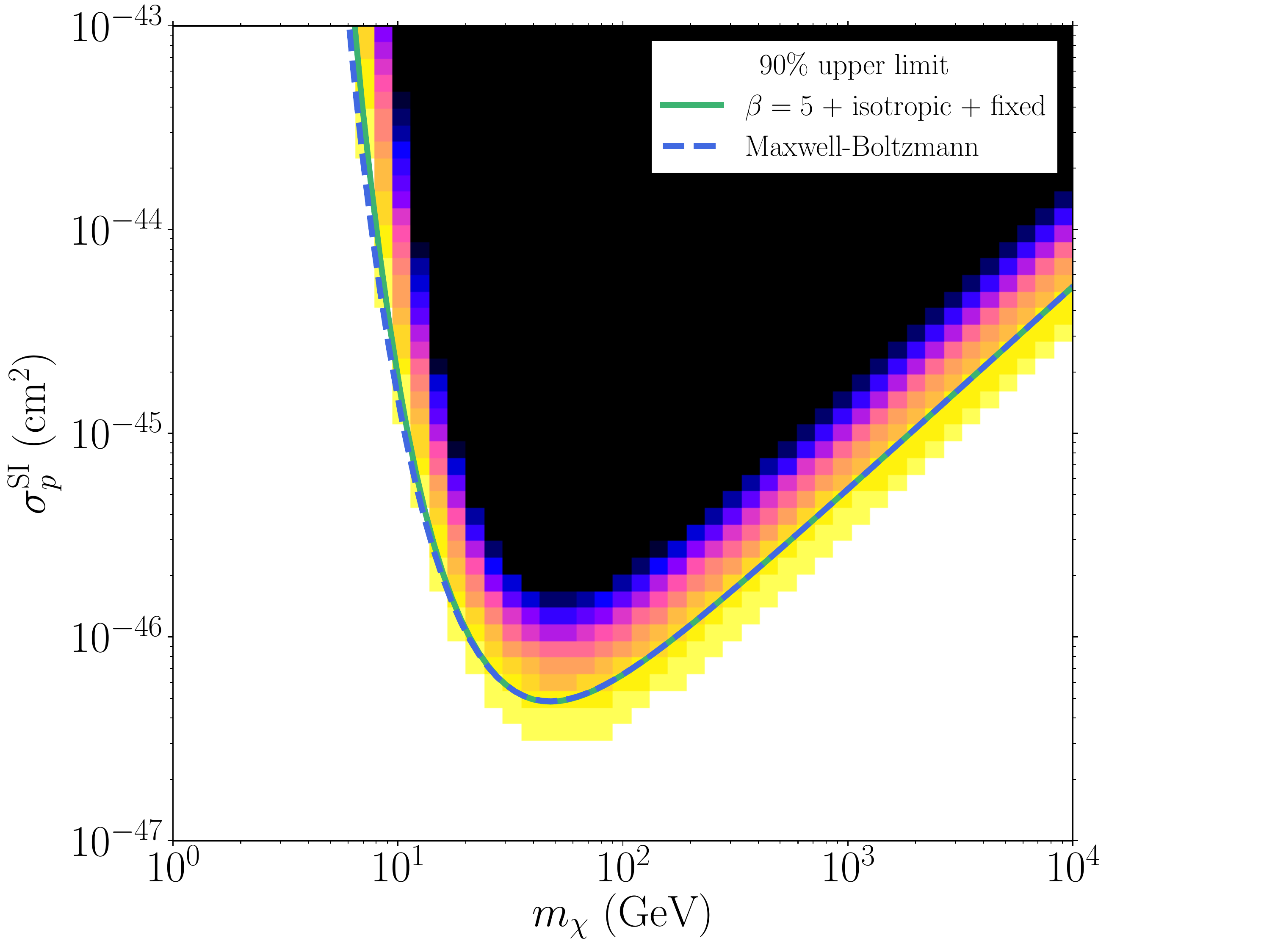}
        \caption{$\beta = 5$}
        \label{fig:isotropic_c}
    \end{subfigure}
    \hspace{-0.75cm}
    \begin{subfigure}[t]{0.48\textwidth}
        \centering
        \includegraphics[height=0.26\textheight]{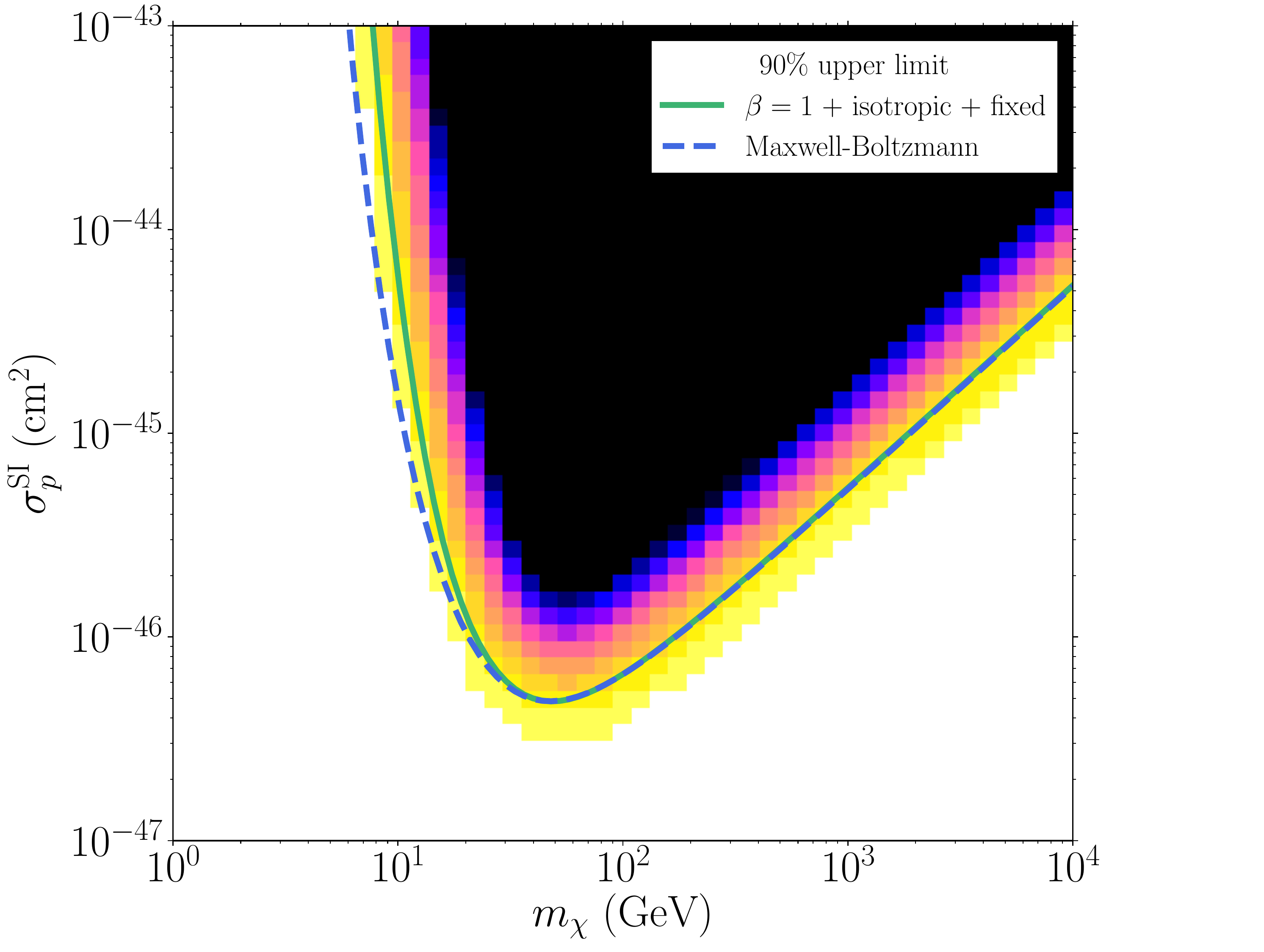}
        \caption{$\beta = 1$}
        \label{fig:isotropic_d}
    \end{subfigure}
    \caption{Chi-squared from \xenon (2018) on the $(m_\chi, \sigsip)$ plane for four values of the parameter, $\beta$, assuming an isotropic velocity distribution. We show the $90\%$ limit assuming an entropic prior (solid green) and that from a Maxwell-Boltzmann (dashed blue).}
    \label{fig:isotropic}
\end{figure}

We begin by assuming an isotropic velocity distribution in the galactic frame. To investigate the dependence of DD searches on the velocity distribution, we marginalise possible departures from a Maxwell-Boltzmann. The hyperparameter $\beta$ governs the strength of our belief in a Maxwellian distribution. As we relax the Maxwellian by decreasing $\beta$, the limit could, a priori, weaken or strengthen.
In \reffig{fig:isotropic} we show maps of
\begin{equation}
\Delta\chi^2 \equiv -2 \ln \frac{\mean{\like}}{\max\limits_{m_\chi, \sigma} \mean{\like}}
\end{equation}
where the average likelihood, $\mean{\like}$, is a function of the hyperparameter, $\beta$, and the DM mass and cross section. We calculate $90\%$ limits from a hybrid approach~\cite{COUSINS1992331} at $\Delta\chi^2 \simeq 1.64$, following \refcite{Workgroup:2017lvb,Athron:2018hpc}. Although we could calculate credible regions in a completely Bayesian approach, we note that hybrid approaches are common in experimental searches and could be adopted by DD experiments themselves.

For $\beta = 100$ in \reffig{fig:isotropic_a}, we see that the $90\%$ limit approximately matches that from a Maxwell-Boltzmann, i.e., at $\beta=100$ we find that we are not sensitive to departures from a Maxwellian. As we decrease our belief in a Maxwellian distribution to $\beta=10$ and $\beta=5$ in \reffig{fig:isotropic_b} and \reffig{fig:isotropic_c}, we see that the \xenon limit becomes slightly weaker than that from a Maxwell-Boltzmann, particularly for DM masses $m_\chi \lesssim 60\gev$. As shown in \reffig{fig:events}, the angle-averaged events function for light DM exhibits a sharp increase in the tail of the Maxwellian distribution and thus \xenon is particularly sensitive to the velocity distribution for light DM. Once we relax to $\beta=1$, \reffig{fig:isotropic_d}, we find pronounced differences for light DM, $m_\chi \lesssim 60\gev$. For heavy DM, $m_\chi \gtrsim 60\gev$, however, the limit stubbornly remains close to that from a Maxwellian distribution. We can understand this by noting that the angle-averaged events function in \reffig{fig:events} is quite flat for DM masses greater than about $60\gev$. As the events function is approximately constant in velocity, we are not sensitive to the velocity distribution.

It is somewhat inevitable that there exists a DM mass at which the angle-averaged events function is approximately flat. For light DM by kinematics we see a sharp rise in the events function near the escape velocity. For heavy DM, high velocities are mildly suppressed. Thus, we find an approximately flat distribution in the transition between these regimes. The fact that this occurs at about $60\gev$ is interesting, as it is close to peak sensitivity and approximately corresponds to $m_\chi \simeq m_h / 2$. Thus, DD limits on WIMPs that annihilate through an on-shell Higgs boson are particularly robust with respect to the velocity distribution.

\subsection{Anisotropic velocity distribution}

\begin{figure}[t]
    \centering
    \begin{subfigure}[t]{0.48\textwidth}
        \centering
        \includegraphics[height=0.26\textheight]{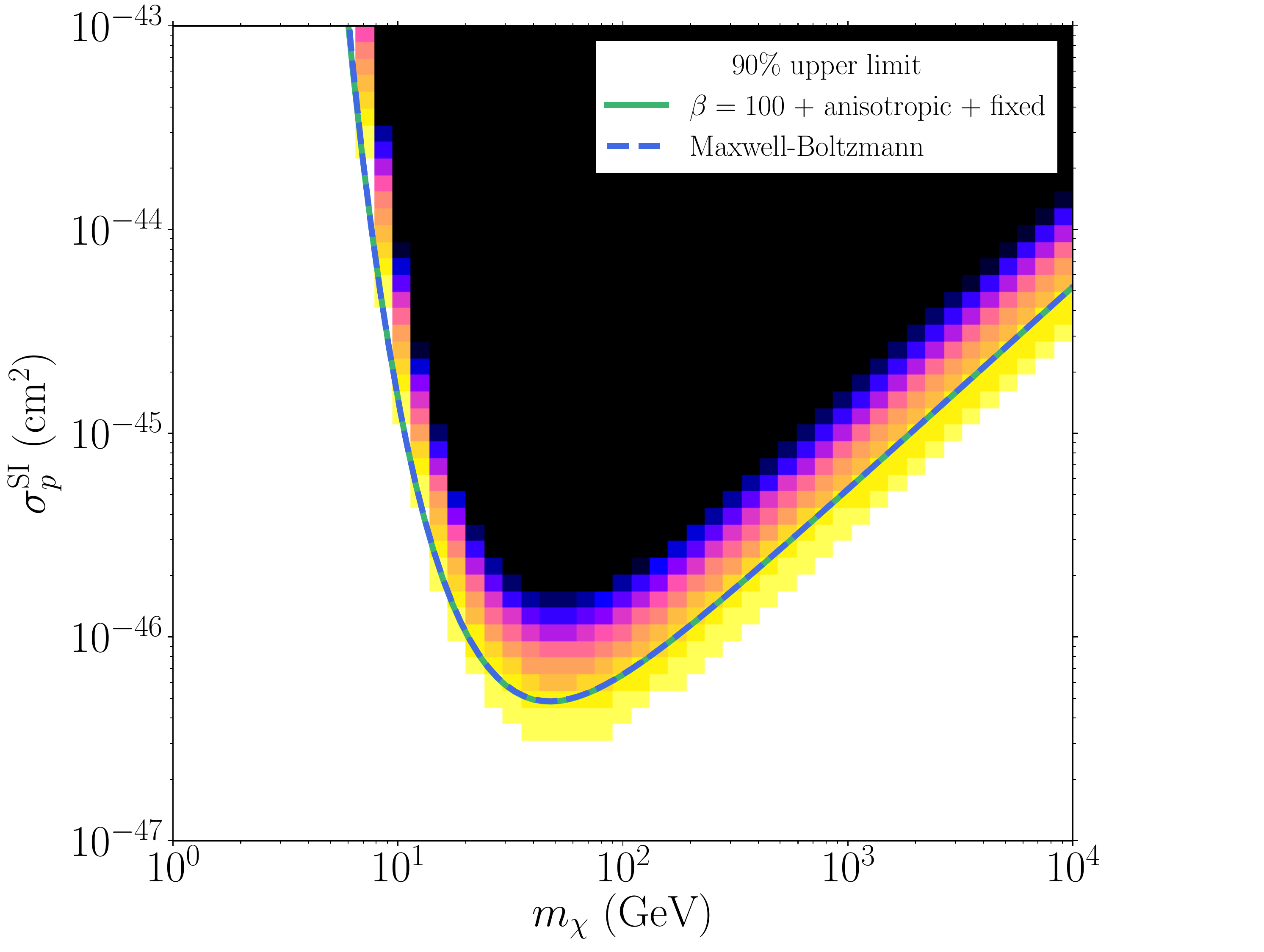}
        \caption{$\beta = 100$}
        \label{fig:anisotropic_a}
    \end{subfigure}
    \hspace{-0.75cm}
    \begin{subfigure}[t]{0.48\textwidth}
        \centering
        \includegraphics[height=0.26\textheight]{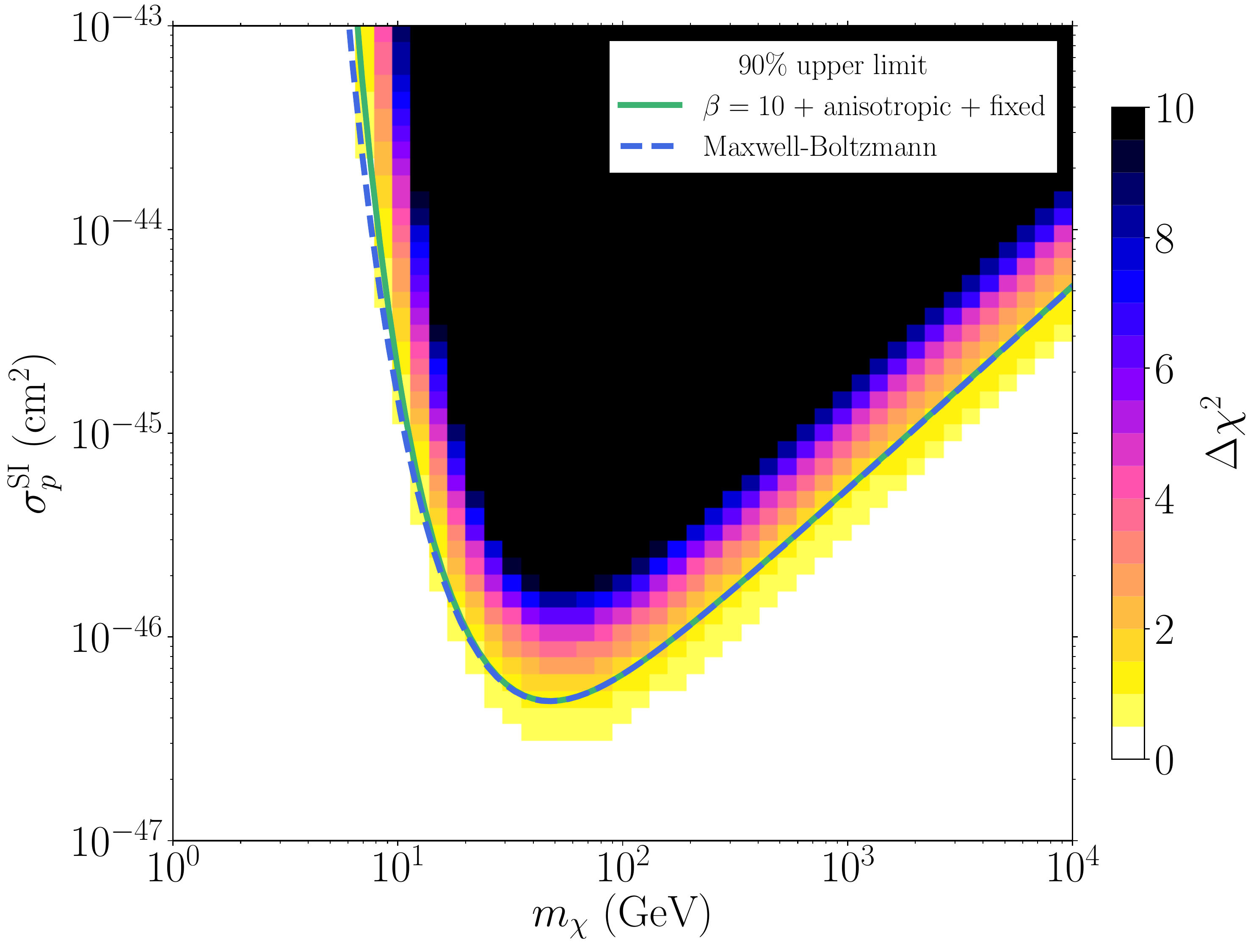}
        \caption{$\beta = 10$}
        \label{fig:anisotropic_b}
    \end{subfigure}

    \begin{subfigure}[t]{0.48\textwidth}
        \centering
        \includegraphics[height=0.26\textheight]{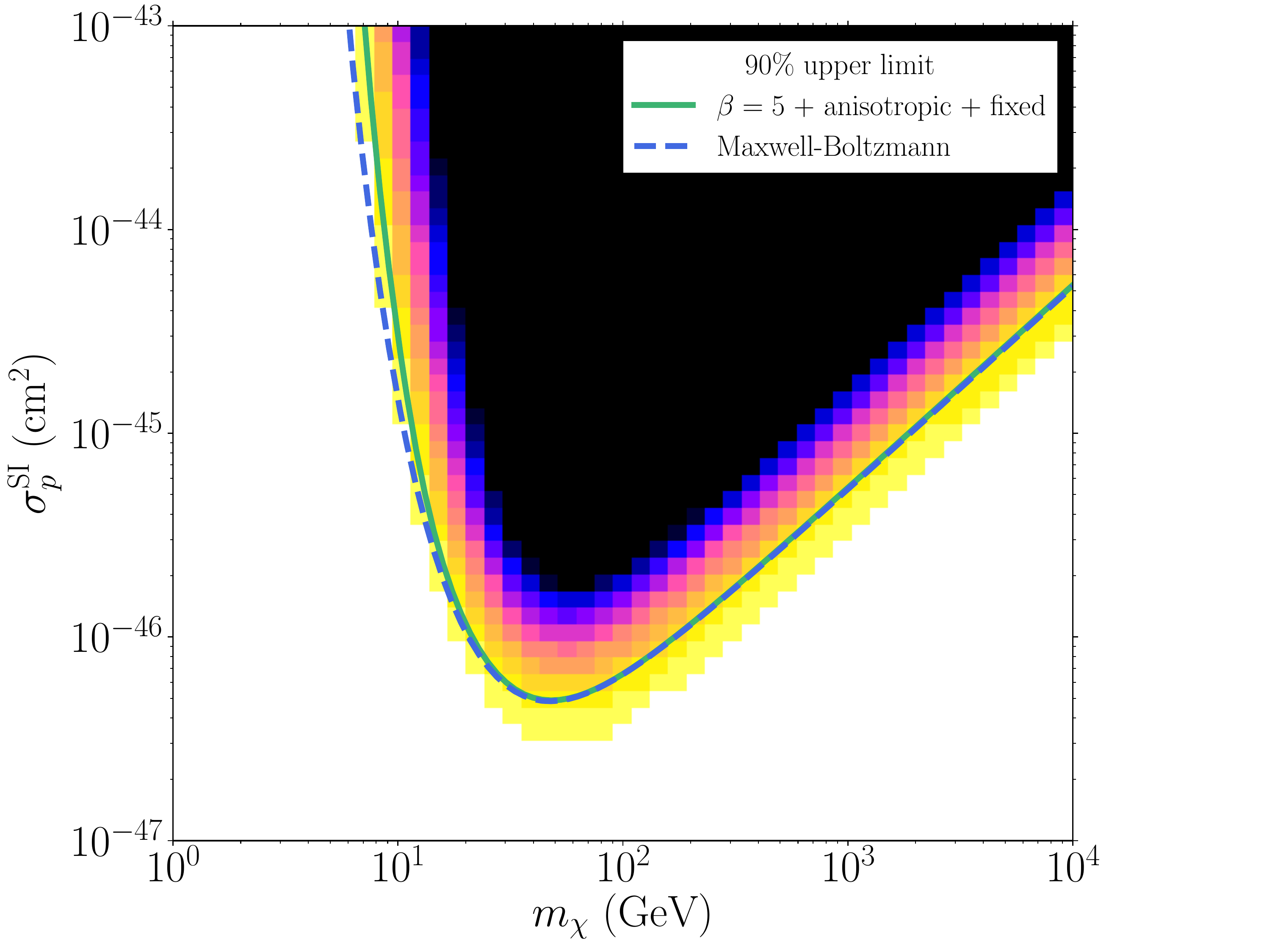}
        \caption{$\beta = 5$}
        \label{fig:anisotropic_c}
    \end{subfigure}
    \hspace{-0.75cm}
    \begin{subfigure}[t]{0.48\textwidth}
        \centering
        \includegraphics[height=0.26\textheight]{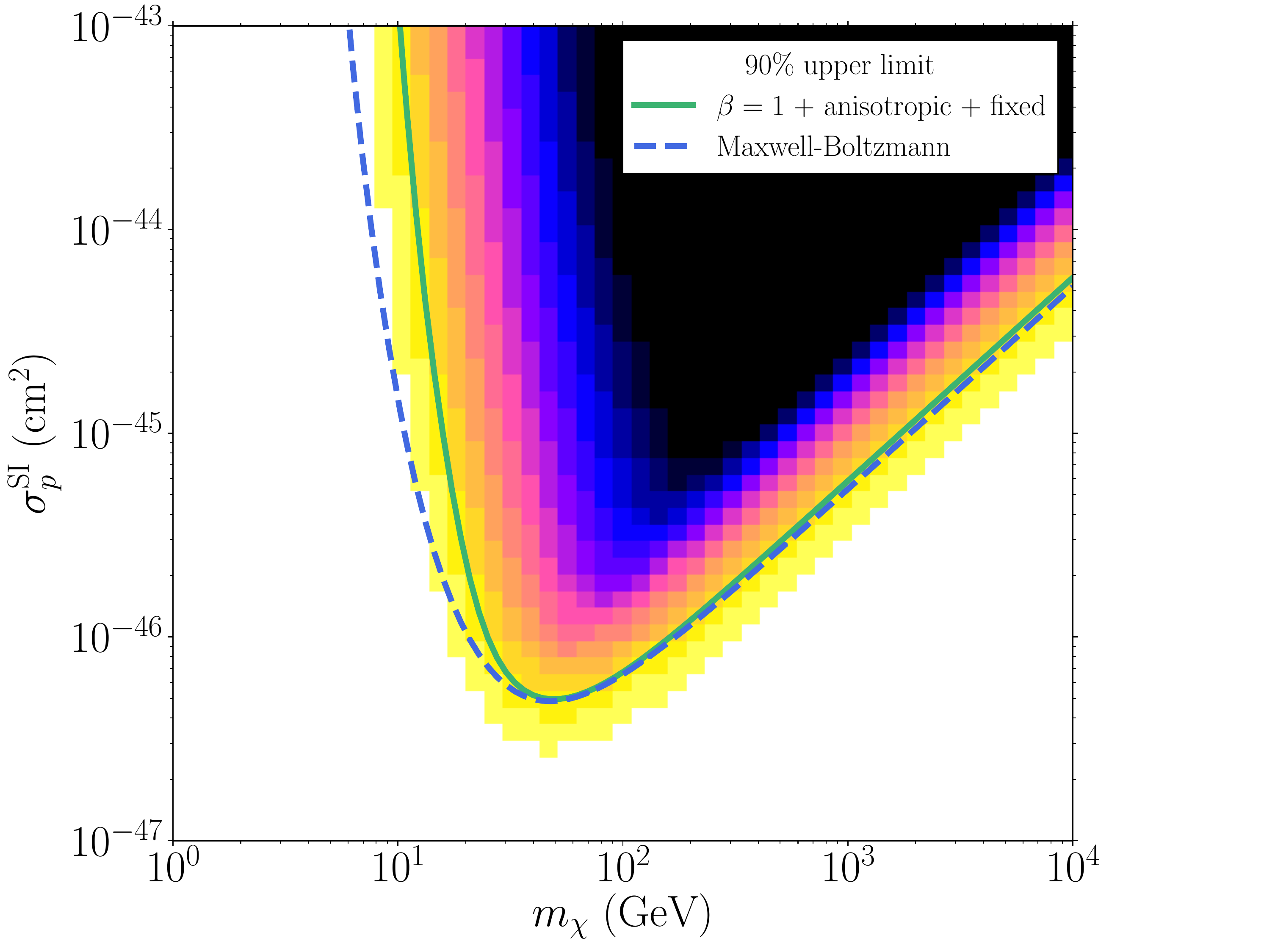}
        \caption{$\beta = 1$}
        \label{fig:anisotropic_d}
    \end{subfigure}
    \caption{Chi-squared from \xenon (2018) on the $(m_\chi, \sigsip)$ plane for four values of the parameter, $\beta$, permitting anisotropic
    departures from a Maxwellian. We show the $90\%$ limit assuming an entropic prior (solid green) and that from a Maxwell-Boltzmann (dashed blue).}
    \label{fig:anisotropic}
\end{figure}

We now relax our assumption of isotropy and place our prior on the magnitude and angular components of the velocity distribution, $\cos\theta$ and $\phi$. This permits anisotropic departures from a Maxwellian distribution. This is important since anisotropy in the galactic frame could be tuned such that the flux of DM particles is zero in the laboratory frame, lifting the limit altogether. In \reffig{fig:anisotropic} we show the $90\%$ limit for four choices of $\beta$.
As expected and as in the isotropic case, for $\beta=100$ in \reffig{fig:anisotropic_a} we find that the limit is approximately that from an isotropic Maxwellian. As we decrease to $\beta=10$ and $\beta=5$ in \reffig{fig:anisotropic_b} and \reffig{fig:anisotropic_c}, we see that the limit weakens. The weakening, although more pronounced than in the isotropic case, remains limited. Even once we relax to $\beta=1$, the weakening is modest, and for DM masses $m_\chi \gtrsim 60\gev$, stubbornly remains close to the Maxwellian limit, as in the isotropic case.

\begin{figure}[t]
    \begin{subfigure}[t]{0.48\textwidth}
        \centering
        \includegraphics[height=0.26\textheight]{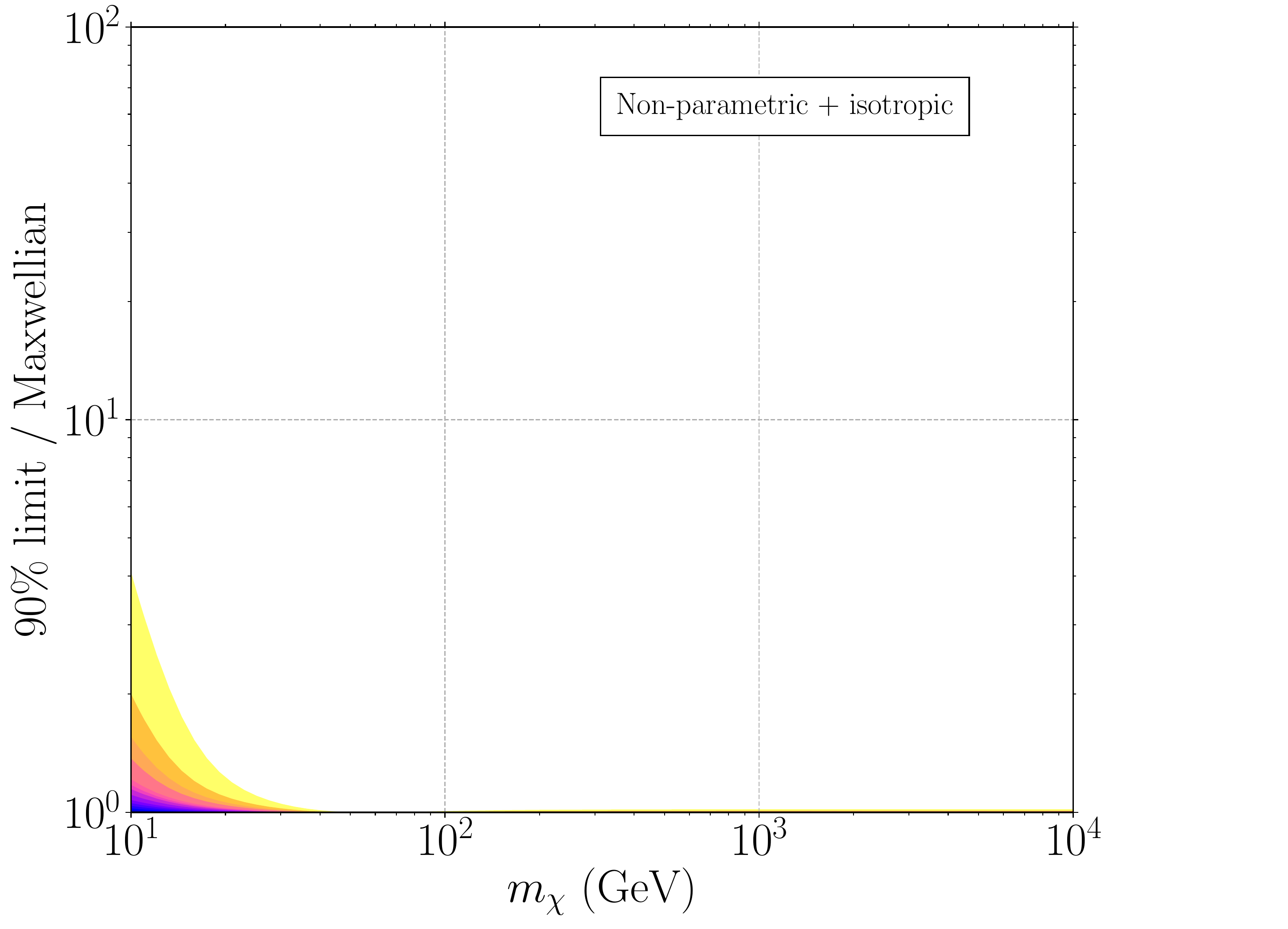}
        \caption{Isotropic}
        \label{fig:impact_isotropic}
    \end{subfigure}
    \begin{subfigure}[t]{0.48\textwidth}
        \centering
        \includegraphics[height=0.26\textheight]{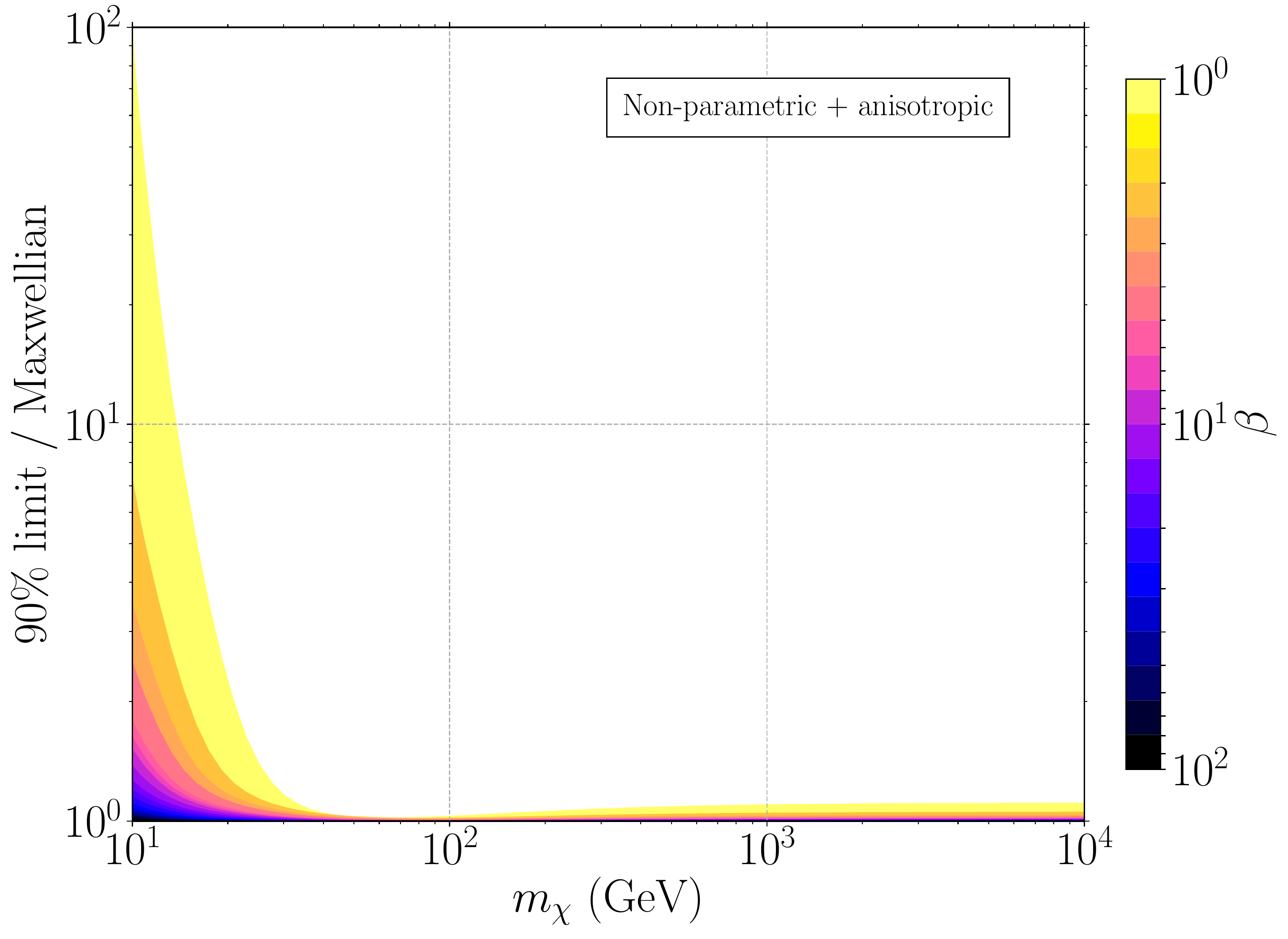}
        \caption{Anisotropic}
        \label{fig:impact_anisotropic}
    \end{subfigure}
    \caption{Change in the $90\%$ limit, relative to the limit from a Maxwellian, from non-parametric uncertainties in the velocity distribution. The hyperparameter $\beta$ governs the strength of our belief in a Maxwell-Boltzmann. We \refsubfig{fig:impact_isotropic} assume an isotropic velocity distribution and \refsubfig{fig:impact_anisotropic} include non-parametric uncertainties in the angular dependence of the velocity distribution.}
    \label{fig:impact}
\end{figure}

In \reffig{fig:impact} we compare our isotropic and anisotropic limits by showing the changes in the limit as we relax the Maxwellian side by side. We divide the limit with that from a Maxwellian. We see that the impact of uncertainty in the velocity distribution in the isotropic case, \reffig{fig:impact_isotropic}, is mild, as it only substantially weakens the limit for light DM and once almost all information about the distribution is disregarded, $\beta \simeq 1$. Even in the most extreme cases, the limit is weakened by less than an order of magnitude. The anisotropic case, in \reffig{fig:impact_anisotropic}, on the other hand, is slightly more dramatic, with noticeable weakening by up to two orders of magnitude for $\beta = 1$. Nevertheless, for DM masses greater than about $60\gev$ the limit stays similar to that from a Maxwell-Boltzmann. We do not investigate DM masses less than $10\gev$, as the limit is acutely sensitive to precision in the tiny efficiency at low recoil energies. We note, though, that in all cases the limit weakens; a priori, it could have strengthened.

\subsection{Parametric uncertainties}

\begin{figure}[t]
    \begin{subfigure}[t]{0.48\textwidth}
        \centering
        \includegraphics[height=0.26\textheight]{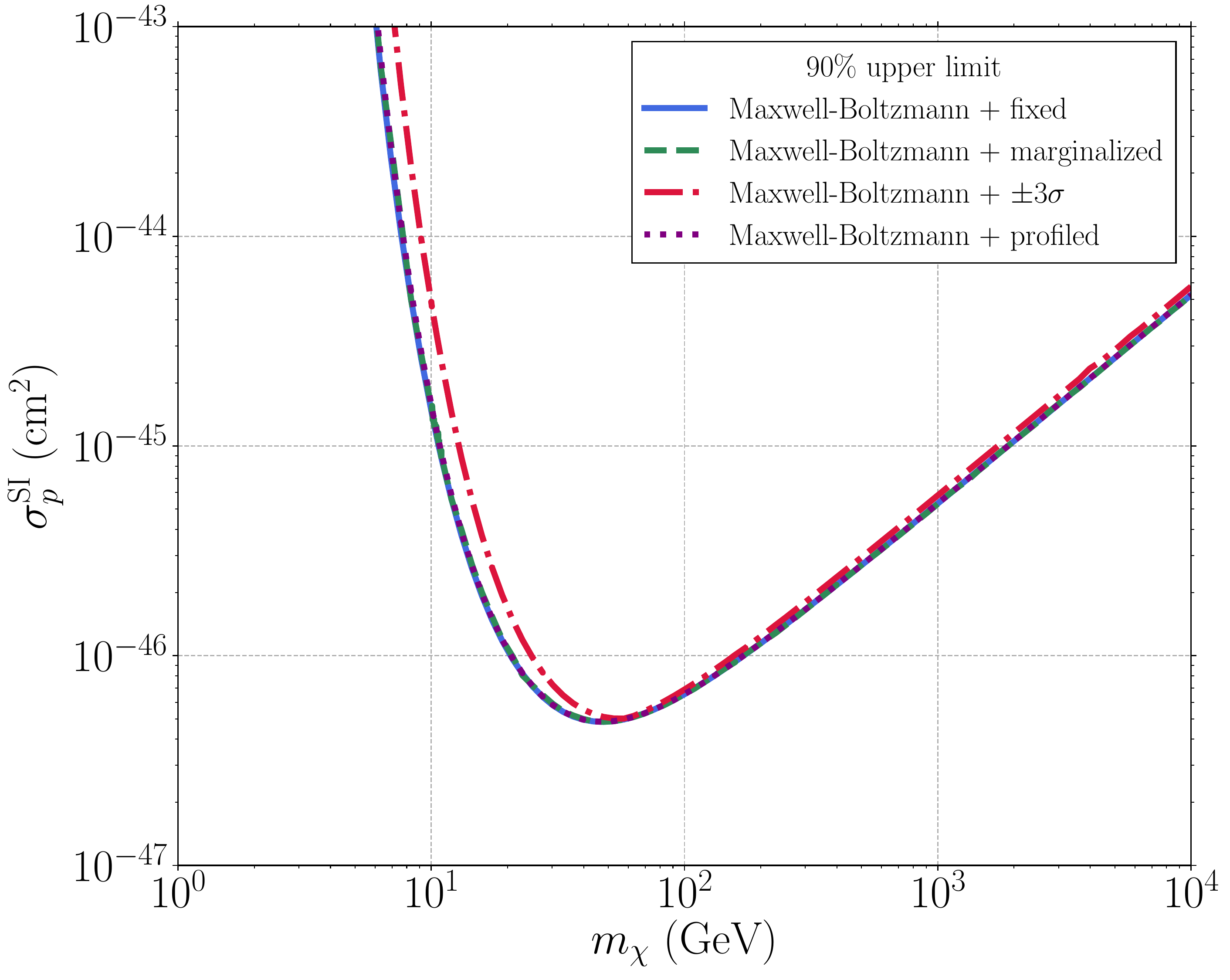}
        \caption{Maxwellian}
        \label{fig:shape_maxwellian}
    \end{subfigure}
    \begin{subfigure}[t]{0.48\textwidth}
        \centering
        \includegraphics[height=0.26\textheight]{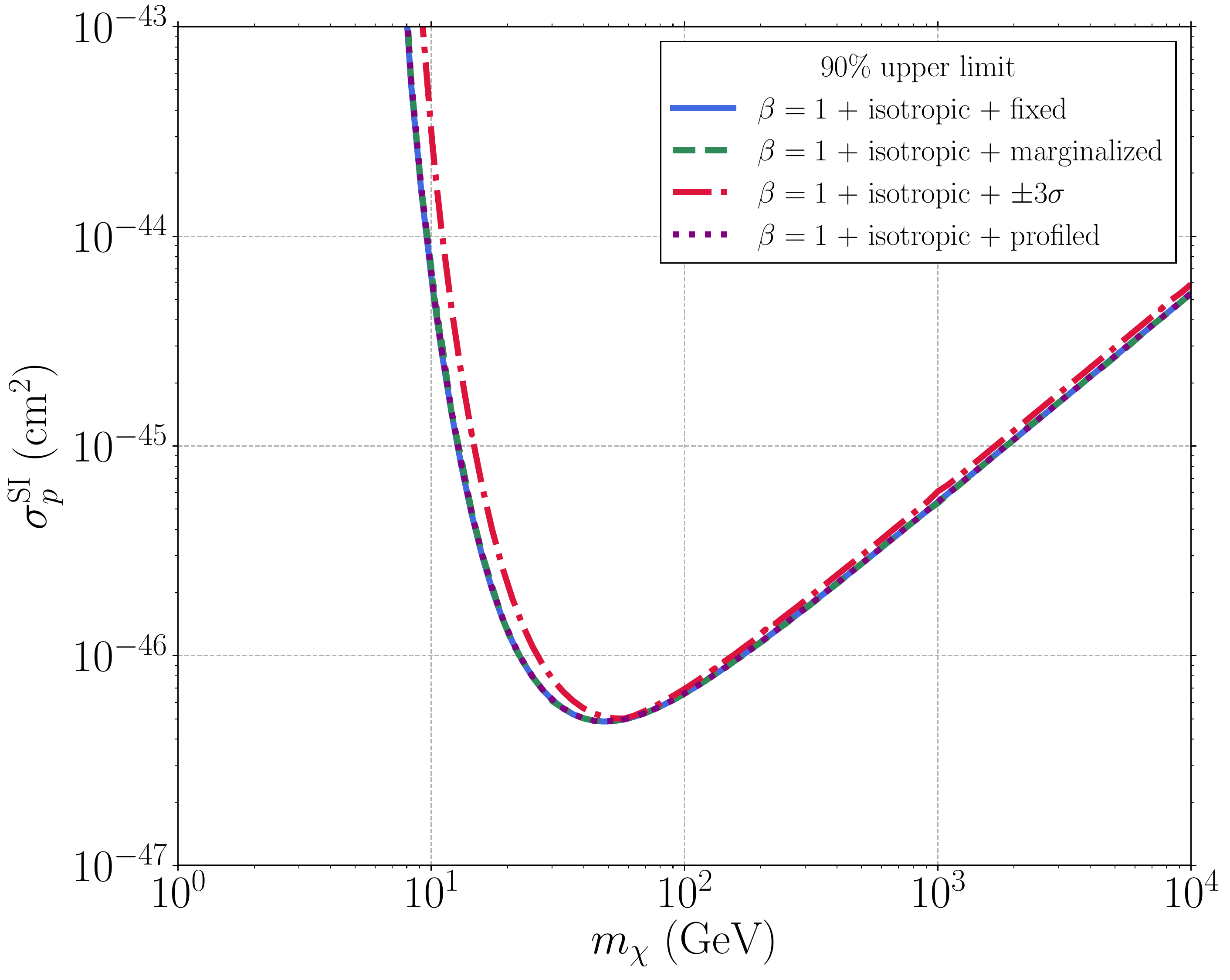}
        \caption{$\beta=1$}
        \label{fig:shape_relaxed}
    \end{subfigure}
    \caption{The $90\%$ limit from \xenon with parametric uncertainties in the velocity distribution. We show $90\%$ limits with the modal and escape velocities fixed (solid blue), averaged (dashed green) and profiled (dotted red). We show the impact on  \refsubfig{fig:shape_maxwellian} a Maxwellian distribution and  \refsubfig{fig:shape_relaxed} with parametric uncertainties from our entropic prior.}
    \label{fig:shape}
\end{figure}

Finally, we consider the impact of parametric uncertainties in the shape parameters of the default distribution. We suspect that the modal and escape velocities are approximately $v_0 = 235 \pm 20 \kms$ and  $v_\text{esc} = 550 \pm 35 \kms$~\cite{Workgroup:2017lvb}. We treat them in four ways: we fix them to their central values; marginalise Gaussian uncertainties in them; permit them to vary by as much as $3\sigma$; and profile Gaussian uncertainties in them. We find, as expected, that their impact is extremely limited. In \reffig{fig:shape} we show the $90\%$ limits from our four treatments. For a Maxwellian, \reffig{fig:shape_maxwellian}, the limits from fixing, marginalising and profiling are extremely similar. The impact of parametric uncertainties is noticeable only when they are permitted to vary by $3\sigma$ without any penalty. The story for the relaxed Maxwellian, \reffig{fig:shape_relaxed}, is similar; the limit cannot be significantly changed by parametric uncertainties.

\section{Discussion and conclusions}\label{sec:conclusions}

We presented a new technique for treating non-parametric uncertainties that applies to any
counting experiment for which the expected number of events may be written as an expectation, e.g., a counting experiment at a collider experiment where the number of events depends upon integrating over
a parton distribution function. We treated non-parametric uncertainties with an multinomial prior that contained a hyperparameter, $\beta$, which governed the
strength of our conviction in a particular parametric model. We detail our result in \refapp{app:average} and \refapp{app:proof}. We
briefly mentioned in \refapp{app:multiple} that our result could generalise to multiple independent counting experiments, though leave a detailed discussion and example to a future work. Our prior was motivated by quantified maximum entropy; but unlike it, it did not suffer from problems with the law of large numbers in the continuum limit. The prior, however, quantized probabilities in multiples of $1/\beta$. It may be desirable to marginalize a prior that permits arbitrary probabilities.

We applied our technique to limits on the scattering cross section of DM from the \xenon experiment. We validated our model of \xenon by reproducing the limit with a Maxwell-Boltzmann distribution. Once we relaxed that distribution, we found only a mild impact from non-parametric uncertainties in the velocity distribution of DM. The impact was greatest when non-parametric uncertainty was included in the angular dependence of the velocity distribution, i.e., in the anisotropic case. For $\beta = 1$ and DM masses less than about $60\gev$, non-parametric uncertainties weakened the upper limit by about two orders of magnitude. Assuming isotropy, however, it weakened by less than one order of magnitude. For DM masses greater than about $60\gev$ the weakening was always mild and the limit at about $60\gev$ was particularly robust with respect to the velocity distribution, as we found that for that mass the events function was approximately flat. The non-parametric uncertainties were, however, significantly greater than the parametric ones. Indeed, after marginalising or profiling them, the impact from uncertainties in the modal and escape velocities was negligible.


Our approach is somewhat in contrast with \refcite{Ibarra:2018yxq}; whereas we constructed a multinomial prior upon velocity distributions and marginalised it, \refcite{Ibarra:2018yxq} selected the most extreme distributions from a set. The former reflects our Bayesian treatment of uncertainty; the latter a frequentist approach. Whilst our results appear to be consistent with \refcite{Ibarra:2018yxq}, in that limits from our marginalised likelihood appear to lie between the extremes found in \refcite{Ibarra:2018yxq}, our results suggest that the impact of uncertainty in the velocity distribution is mild.\footnote{Note, however, that \refcite{Ibarra:2018yxq} used 2017 \xenon results \cite{Aprile:2017iyp}, in which no events were observed.} The formalism itself should lend itself to inclusion in global fits of DM models, as it is not especially computationally demanding. We briefly describe our publicly available implementation in \refapp{app:code}. The ordinary treatment of DD experiments requires a single integral upon the velocity distribution; this one requires an integral for every observed event. Previously global fits of DM models, e.g., \refcite{Balazs:2017ple,Athron:2018ipf}, included at most parametric uncertainties. We can now, however, incorporate coherently all major sources of uncertainty in DD experiments in official limits and global fits.

\bibliographystyle{JHEP}
\bibliography{max_ent_2,max_ent_1}

\appendix

\section{Averaged Poisson likelihood}\label{app:average}

We discretize the velocity distribution by dividing the velocity into $r$ bins of volume $\Delta v^3$, but ultimately we take a continuum limit. We imagine a team of monkeys throwing $\beta$ balls into $r$ bins according to the probabilities in the default model, $m_i \equiv m(v_i) \Delta v^3$. We denote the number of balls in each bin by $n_i$. This is a multinomial process; the probabilities of different occupation numbers, $n_i$, are
\begin{equation}\label{eq:discrete}
\Pg{\vec n}{\vec m} =
\begin{cases}
      \beta! \prod_i \frac{m_i^{n_i}}{n_i!} & \text{if $\sum_i n_i = \beta$}\\
      0 & \text{otherwise}
\end{cases},
\end{equation}
where $\vec n$ denotes the set $\{n_1, n_2, \dots, n_r\}$ and similarly, $\vec m$ represents our default model, a Maxwellian.
For two bins this is a binomial distribution. All products and sums are, unless otherwise specified, over all $r$ bins. This process generates normalised velocity distributions $f_i = n_i / \beta$. By making Stirling approximations for the factorials in \refeq{eq:discrete}, we find that
\begin{equation}\label{eq:approx_QME}
\Pg{\vec f}{\vec m} \propto e^{\beta S[\vec f, \vec m]},
\end{equation}
where $\vec f$ represents a discretized velocity distribution. Thus a prior based on a multinomial resembles the QME prior in \refeq{eq:QME}; in fact, the latter is an analytic continuation of a multinomial. As discussed in \refsec{sec:recap}, the QME prior is not divisible, leading to a dependence on the parameterization and binning, and problems in the continuum limit. The problems stem from the fact that properties of discrete distributions, e.g., the fact that the sum of two Poisson variables is another Poisson variable with a mean that is summed, are broken by analytic continuation.

Returning to the average Poisson likelihood, we may write
\begin{align}
\mean{\like} &= \sum_{\vec f} \frac{e^{-\lambda} \lambda^q}{q!} \cdot \pg{\vec f}{\vec m}\\
\label{eq:integral_multinomial}
& = \sum_{\vec n} \frac{e^{-\lambda} \lambda^q}{q!} \cdot \beta! \prod_i \frac{m_i^{n_i}}{n_i!}.
\end{align}
We may rewrite the final line by combining the exponential factor with the $m_i^{n_i}$ terms,
\begin{equation}\label{eq:multinomial}
\mean{\like} = \frac{1}{q!} \cdot \left(\sum_i m_i e^{-w_i / \beta }\right)^\beta  \cdot  \beta! \sum_{\vec n} \lambda^q \cdot \prod_i \frac{m_i^{\prime n_i}}{n_i!},
\end{equation}
where we defined the modified probabilities
\begin{equation}
m^\prime_i = \frac{m_i e^{-w_i / \beta }}{\sum_j m_j e^{-w_j / \beta }}.
\end{equation}
The sum in \refeq{eq:multinomial} is equivalent to the expectation of $\lambda^q = \left(\sum w_i n_i / \beta \right)^q$, where $n_i$ follows a multinomial distribution with $\beta$ trials and event probabilities $m_i^\prime$. That is,
\begin{equation}
\mean{\like} =  \frac{1}{q!} \cdot \left(\sum_i m_i e^{-w_i / \beta }\right)^\beta  \cdot \mean{\left(\sum_i w_i n_i / \beta\right)^q},
\end{equation}
where $n_i \sim M(\beta, \vec m^\prime)$. We compute this expectation in \refapp{app:proof}, finding that ultimately it can be written in a form involving only expectations under the default distribution. For $q=2$, for example, we find that
\begin{equation}\label{eq:result}
\mean{\like} =
\frac{1}{2}
\mean{e^{-w / \beta}}_m^\beta
\left(
\frac{\beta - 1}{\beta} \frac{\largemean{w e^{-w / \beta}}_m^2}{\largemean{e^{-w / \beta}}_m^2} + \frac{1}{\beta} \frac{\largemean{w^2 e^{-w / \beta}}_m}{\largemean{e^{-w / \beta}}_m}
\right),
\end{equation}
where $\mean{y}_m \equiv \sum y_i m_i$. We may now take the continuum limit of the discrete velocity distributions, replacing $\mean{y}_m \to \int  y(\vec v) \cdot m(\vec v) \dif^3 v$.

\section{Expectation from multinomial}\label{app:proof}

We wish to find the expectation of $\lambda^q = \left(\sum w_i n_i / \beta \right)^q$, where $n_i$ follow a multinomial distribution with $\beta$ trials and event probabilities $m_i$.
First, we utilise a property of the factorial moments of multinomials~\cite{doi:10.1093/biomet/49.1-2.65},
\begin{equation}
\mean{\prod_j (n_j)_{(a_j)}} = \beta_{(A)} \prod_i m_i^{a_i}, 
\end{equation}
where $A \equiv \sum a_i$ and $\beta_{(k)}$ denotes the falling factorial. We turn this into an expression for the moments using Stirling numbers,
\begin{equation}
\mean{\prod_j n_j^{p_j}} = \sum_{\vec a} \stirling{\vec p}{\vec a} \beta_{(A)} \prod_i m_i^{a_i},  
\end{equation}
where our notation is that we write a product of Stirling numbers as,
\begin{equation}
\stirling{\vec p}{\vec a} \equiv \prod_i \stirling{p_i}{a_i},
\end{equation}
where $\stirling{s}{t}$ is a Stirling number. The sum is from $0$ to $p$ for every power.

Second, we use the multinomial theorem~\cite{multinomial_theorem} to write,
\begin{equation}
\left(\sum w_i n_i / \beta\right)^q = \frac1{\beta^q} \sum_{\vec p} \binom{q}{\vec p} \prod_i w_i^{p_i} \prod_i n_i^{p_i},    
\end{equation}
where the multinomial coefficient $\binom{q}{\vec p} \equiv q! /  \prod p_i!$ and we sum upon $\vec p$ subject to the constraint that $\sum p_i = q$. Combining, we have
\begin{align}
\mean{\left(\sum w_i n_i / \beta \right)^q} &= \frac1{\beta^q} \sum_{\vec p} \binom{q}{\vec p} \prod_i w_i^{p_i} \mean{\prod_i n_i^{p_i}}\\    
&= \frac{1}{\beta^q} \sum_{\vec p} \binom{q}{\vec p} \prod_j w_j^{p_j}  \sum_{\vec a} \stirling{\vec p}{\vec a} \beta_{(A)} \prod_j m_j^{a_j}.  
\end{align}
This equals
\begin{equation}\label{eq:general_result}
\mean{\left(\sum w_i n_i / \beta \right)^q}=  \sum_{\vec p} \frac{\beta_{(k)}}{\beta^q} \frac{1}{F} \binom{q}{\vec p} \prod_j \sum_i w^{p_j}_i m_i =  \sum_{\vec p} \frac{\beta_{(k)}}{\beta^q}  \frac{1}{F}  \binom{q}{\vec p} \prod_j \mean{w^{p_j}}_m
\end{equation}
where $k$ is the number of non-zero powers in $\vec p$; and $F = \prod_i c_i!$, where $c_i$ is the number of times a power appears, is a factorial factor that accounts for the cases in which the powers are not unique. We define $\mean{y}_m \equiv \sum_i y_i m_i$, though note that we may take the continuum limit throughout $\mean{y}_m \to \int y(\vec v) \cdot m(\vec v) \dif^3 v$. We plan to present a more detailed proof in a future work dedicated to this result.

The number of terms in the sum is equal to the number of unique ways of partitioning $q$ into any number of smaller terms, i.e., the partition function of $q$. The first ten partitions are $1$, $2$, $3$, $5$, $7$, $11$, $15$, $22$, $30$ and $42$, and, asymptotically, the number of partitions grows exponentially with $q$.  The formula allows us to express complicated moments of a multinomial in terms of simpler ones; in this regard, it is similar to Wick's theorem for Gaussian moments. As the formula is complicated, however, we illustrate it for $q=1$, $2$ and $3$:
\begin{description}
\item[$q=0$] This case is trivial, $\mean{\left(\sum w_i n_i / \beta\right)^0} = \langle 1 \rangle = 1$.
\item[$q=1$] There is a single partition of $q$ with $\vec p= \{1\}$. Thus, $F=1$, $k=1$, $\binom{q}{\vec p} = 1$, and  
we find 
\begin{equation}
\mean{\left(\sum w_i n_i / \beta\right)} = \sum_i w_i m_i \equiv \mean{w}_m.  
\end{equation}

\item[$q=2$] There are two partitions of $q$ with $\vec p= \{1 + 1, 2\}$. Thus there are two terms in our result. 

For the first term, $1 + 1$, we find $k=2$, since there are two non-zero powers; $F=2!$, since a power is repeated twice; and
$\binom{q}{\vec p} = 2$.  Thus this term is
\begin{equation}
\frac{\beta_{(2)}}{\beta^2} \mean{w}_m^2 = \frac{\beta - 1}{\beta}  \mean{w}_m^2.
\end{equation}
For the second term, $2$, we find $k=1$, $F=1$, and $\binom{q}{\vec p} = 1$. Thus this term is
\begin{equation}
\frac{\beta_{(1)}}{\beta^2} \mean{w^2}_m   = \frac{1}{\beta} \mean{w^2}_m.
\end{equation}
Thus summing the two terms we find,
\begin{equation}
\mean{\left(\sum w_i n_i / \beta\right)^2}   =  \frac{1}{\beta} \largemean{w^2}_m +  \frac{\beta - 1}{\beta}  \largemean{w}_m^2.
\end{equation}

\item[$q=3$] There are three partitions of $q$ with $\vec p= \{1 + 1 + 1, 1 + 2, 3\}$. Thus there are three terms in our result. 

For the first term, $1 + 1 + 1$, we find $k=3$, since there are three non-zero powers; $F = 3! = 6$, since a power is repeated three times; and
$\binom{q}{\vec p} = 6$. Thus this term is
\begin{equation}
\frac{\beta_{(3)}}{\beta^3} \mean{w}_m^2 = \frac{(\beta - 1)(\beta - 2)}{\beta^2} \mean{w}_m^3.
\end{equation}
For the second term, $1 + 2$, we find $k=2$, $F=1$, and $\binom{q}{\vec p} = 3$. Thus this term is
\begin{equation}
3 \frac{\beta_{(2)}}{\beta^3} \largemean{w}_m \largemean{w^2}_m = 3 \frac{(\beta - 1)}{\beta^2} \largemean{w}_m \largemean{w^2}_m.
\end{equation}
Finally, for the third term, $3$, we find $k=1$, $F=1$, and $\binom{q}{\vec p} = 1$. Thus this term is
\begin{equation}
\frac{\beta_{(1)}}{\beta^3} \mean{w^3}_m = \frac{1}{\beta^2} \mean{w^3}_m.
\end{equation}
Thus summing the three terms we find,
\begin{equation}\label{eq:q_3_case}
\mean{\left(\sum w_i n_i / \beta\right)^3} = \frac{(\beta - 1)(\beta - 2)}{\beta^2} \largemean{w}_m^3 + 3 \frac{(\beta - 1)}{\beta^2} \largemean{w}_m \largemean{w^2}_m
+ \frac{1}{\beta^2} \largemean{w^3}_m.
\end{equation}
\end{description}

\subsection{Generalized result}\label{app:multiple}

We may in fact generalise our result in \refeq{eq:general_result} to e.g., the expectation of 
\begin{equation}
\lambda_1^{q_1} \lambda_2^{q_2} = \left(\sum (w_1)_i n_i / \beta \right)^{q_1} \left(\sum (w_2)_i n_i / \beta \right)^{q_2},
\end{equation}
where $n_i$ follow a multinomial distribution with $\beta$ trials and event probabilities $m_i$. \refeq{eq:general_result} applies to the special case in which $w_1 = w_2$. Such a 
result would be useful for e.g., marginalizing the uncertainty in a likelihood from two or more independent counting experiments. In this case, we may in fact apply
\refeq{eq:general_result} by using $q = q_1 + q_2$ and replacing e.g., $\mean{w^n}_m$ terms with appropriate generalisations involving $\mean{w_1}_m$ and $\mean{w_2}_m$ etc. For example,
\begin{equation}
\begin{split}
\mean{\lambda_1^2 \lambda_2^1} &= \frac{(\beta - 1)(\beta - 2)}{\beta^2} \largemean{w_1}_m^2 \largemean{w_2}_m +
\frac{(\beta - 1)}{\beta^2} \left(\largemean{w_1^2}_m \largemean{w_2}_m + 2 \largemean{w_1 w_2}_m \largemean{w_1}_m \right)\\
& + \frac{1}{\beta^2} \largemean{w_1^2 w_2}_m,   
\end{split}
\end{equation}
which bears a close resemblence to \refeq{eq:q_3_case}. We leave a detailed discussion of this generalised case and an application to the likelihood from multiple independent counting experiments to a future work.

\section{Computer code --- \texorpdfstring{\pyth{veltropy}}{veltropy}}\label{app:code}

We provide a Python module implementing the events function and averaging upon our entropic prior. The module should be downloaded from \url{https://github.com/andrewfowlie/veltropy/archive/master.zip} or cloned by
\begin{python}
git clone https://github.com/andrewfowlie/veltropy
\end{python}
The requirements are listed in \pyth{requirements.txt} and may be installed by
\begin{python}
pip install -r requirements.txt
\end{python}

There are four main classes, which should be imported by
\begin{python}
from veltropy import EventsAtVelocity, Poisson, Relax, Shape
\end{python}
There are many further classes and methods documented in the code, which could be used for arbitrary anisotropic velocity distributions and experiments. See e.g.,
\pyth{experiment.py} for the implementation of XENON1T. There is an example program,
\begin{python}
python limit_example.py
\end{python}
which plots an upper limit on the cross section with a $\beta = 10$ entropic prior about a Maxwellian default distribution.

\subsection{The events class --- \texorpdfstring{\pyth{EventsAtVelocity}}{EventsAtVelocity}}

This class is defined in  \pyth{events.py}. This class builds an events function, $w(\vec v)$, for DM of a particular mass. By default, it uses XENON1T. E.g.,
\begin{python}
mass = 100.  # GeV
w = EventsAtVelocity(mass)
\end{python}
The main methods are the events functions in the galactic frame, e.g.,
\begin{python}
velocity = 200.  # km/s
w(velocity)  # Angle-averaged
cos_theta = 0.
w(velocity, cos_theta)  # Not angle-averaged
w.plot()  # Plots events function
\end{python}
There are further methods for the events function in e.g., the earth frame. We can convolute with a velocity distributions by e.g.,
\begin{python}
from relax import MB

velocity_dist = MB()  # Maxwell-Boltzmann with default parameters
n_events = velocity_dist * w  # This performs integral over velocity
\end{python}

\subsection{Poisson likelihood --- \texorpdfstring{\pyth{Poisson}}{Poisson}}

This class is defined in \pyth{poisson.py} and calculates the likelihood $\like$, as a function of the cross section, assuming a Maxwell-Boltzmann velocity distribution by default. 
We use it by, e.g.,
\begin{python}
mass = 100.  # GeV
w = EventsAtVelocity(mass)
poisson = Poisson(w)
sigma = 1E-42  # cm^2
poisson.loglike(sigma)
\end{python}
This requires an \pyth{EventsAtVelocity} instance. We can optionally specify a velocity distribution, e.g., \pyth{Poisson(w, velocity_dist=my_dist)}. There are further properties for inspecting results, e.g.,
\begin{python}
relax.chi_squared_limit()  # 90
relax.best_fit_sigma   # Best-fit cross section
\end{python}
return the $90\%$ upper limit and best-fit cross section.

\subsection{The relaxed Maxwellian class --- \texorpdfstring{\pyth{Relax}}{Relax}}

This class is defined in \pyth{relax.py} and calculates the likelihood averaged upon an entropic prior, $\mean{\like}$, as a function of the cross section. We use it by, e.g.,
\begin{python}
beta = 10
mass = 100.  # GeV
w = EventsAtVelocity(mass)
relax = Relax(beta , w, isotropic=True)
sigma = 1E-42  # cm^2
relax.loglike(sigma)
\end{python}
This requires the hyperparameter \pyth{beta } and an \pyth{EventsAtVelocity} instance. We can optionally specify a velocity distribution and whether we wish to assume isotropy. The further methods 
are similar to that for \pyth{Poisson}.

\subsection{Averaging upon shape parameters --- \texorpdfstring{\pyth{Shape}}{Shape}}

This class is defined in \pyth{shape.py} and calculates the likelihood averaged upon parametric uncertainties in the modal and escape velocities. The usage is similar to the \pyth{Relax} and  \pyth{Poisson} classes, e.g.,
\begin{python}
beta = 10
mass = 100.  # GeV
w = EventsAtVelocity(mass)
relax = Relax(beta , w, isotropic=True)
shape = Shape(relax)
sigma = 1E-42  # cm^2
shape.loglike(sigma)
\end{python}
Note that we initialise a \pyth{Shape} instance with a \pyth{Relax} or \pyth{Poisson} instance.

\end{document}